\documentclass{vgtc}                          

\graphicspath{{figures/}{pictures/}{images/}{./}} 

\usepackage{times}                     

\usepackage{tabu}                      
\usepackage{booktabs}                  
\usepackage{lipsum}                    
\usepackage{mwe}                       

\usepackage{soul}

\usepackage{mathptmx}                  

\onlineid{0}

\vgtccategory{Research}

\vgtcinsertpkg

\title{Assist, Don’t Analyze: Designing a Lightweight GenAI Interface for Visual Data Analysis}

\author{Ratanond Koonchanok\thanks{e-mail: rkoonch@iu.edu}\\ %
        \scriptsize Indiana University Indianapolis %
\and Alex Kale\thanks{e-mail: kalea@uchicago.edu}\\ %
     \scriptsize University of Chicago %
\and Khairi Reda\thanks{e-mail: redak@uic.edu}\\ %
     \scriptsize University of Illinois Chicago}

\teaser{
  \centering
  \includegraphics[width=\linewidth]{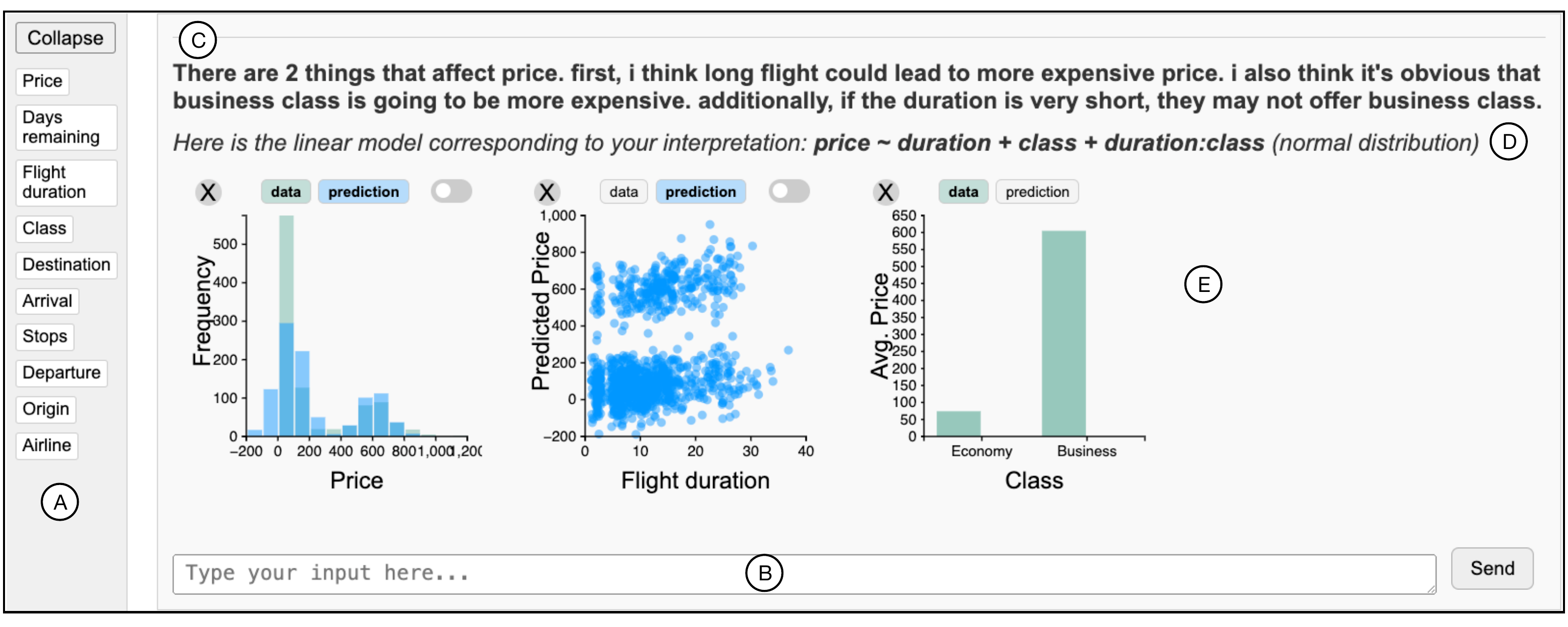}
  \caption{Overview of the interface. The left panel displays the data attributes (A). Users can type their query into the text box (B). After clicking ``Send'', the query appears at the top of the corresponding result panel (C). In this example, the linear model generated from the user input is displayed (D), followed by visualizations corresponding to the variables in the model (E).}
  
  \label{fig:teaser}
}

\abstract{

Recent advances in Generative AI have transformed how users interact with data analysis through natural language interfaces. However, many systems rely too heavily on LLMs, creating risks of hallucination, opaque reasoning, and reduced user control. We present a hybrid visual analysis system that integrates GenAI in a constrained, high-level role to support statistical modeling while preserving transparency and user agency. GenAI translates natural language intent into formal statistical formulations, while interactive visualizations surface model behavior, residual patterns, and hypothesis comparisons to guide iterative exploration. Model fitting, diagnostics, and hypothesis testing are delegated entirely to a structured R-based backend, ensuring correctness, interpretability, and reproducibility. By combining GenAI-assisted intent translation with visualization-driven reasoning, our approach broadens access to modeling tools without compromising rigor. We present an example use case of the tool and discuss challenges and opportunities for future research.

} 


\begin{document}


\firstsection{Introduction}
\maketitle

Generative AI (GenAI) and large language models (LLMs) have transformed how users interact with data. These technologies promise to lower the barrier to statistical analysis by allowing users to express analytical goals in natural language, without needing to write formal code. However, as LLMs are increasingly embedded into end-to-end data science systems, a new set of challenges arises. Systems that delegate full analytical control to GenAI models often compromise on transparency, correctness, and user control. While these models can appear flawless, their tendency to hallucinate, misinterpret user goals, or produce unverifiable outputs makes them unreliable for analytical tasks. These issues are particularly concerning when GenAI is used for deeper forms of analytical reasoning, such as model selection and statistical inference. In such cases, the lack of visibility into the underlying computations can hinder reproducibility and analytical rigor. Users may unknowingly adopt flawed conclusions, assuming the output is trustworthy. This dynamic introduces a new form of risk, analogous to p-hacking, where researchers (intentionally or not) engage in practices that inflate the likelihood of statistically significant findings. With GenAI, this could manifest as ``prompt hacking'' which may include crafting multiple prompts to obtain desired outcomes or failing to validate whether prompts yield consistent results across different models \cite{morris2024prompting}. Therefore, while GenAI holds great promise, its integration into analytical workflows must be approached with caution.

While fully automated GenAI systems can generate statistical models or insights quickly, they often lack transparency and interpretable feedback, making it difficult for users to trust or validate the results. Conversely, traditional visualization tools provide interpretability and fine-grained control but require significant statistical expertise to specify models, diagnose assumptions, and test hypotheses. This creates a gap for users who have domain knowledge but limited modeling experience, as they need accessible ways to express analytic intent while still seeing, understanding, and verifying the analytical process. By integrating GenAI within an interactive visualization environment, our approach leverages the strengths of both. GenAI lowers the barrier to model specification, while visualization provides transparent, iterative feedback that anchors users’ reasoning and preserves their agency and accountability.

In this paper, we present a complementary approach that tightly integrates LLMs with a visual statistical modeling workflow, while intentionally restricting the role of the LLM to a high-level interface layer. Our system enables users to pose analytical questions in natural language, such as specifying a linear model or testing for statistical significance, and pairs these capabilities with interactive visualizations that surface model behavior, residual patterns, and hypothesis comparisons. The LLM serves as a facilitator, interpreting user intent and mapping it to a predefined set of actions before being passed to the R backend. All model fitting, diagnostics, and inference are performed entirely within the R-based backend, ensuring correctness, transparency, and reproducibility. This design preserves the benefits of natural language interaction while avoiding the risks associated with opaque or fully autonomous AI-driven analysis. Through this design, our tool strikes a balance between accessibility and rigor. We demonstrate its utility through an example use case, showing how a non-expert analyst can iteratively construct and refine a statistical model using natural language, visualization-driven feedback, and guided suggestions while retaining full visibility into the underlying methods. Our work contributes to an emerging class of GenAI-integrated visual analytics tools that emphasize reliability, interpretability, and human-in-the-loop control in statistical workflows.







\begin{figure*}
    \centering
    \includegraphics[width=1\linewidth]{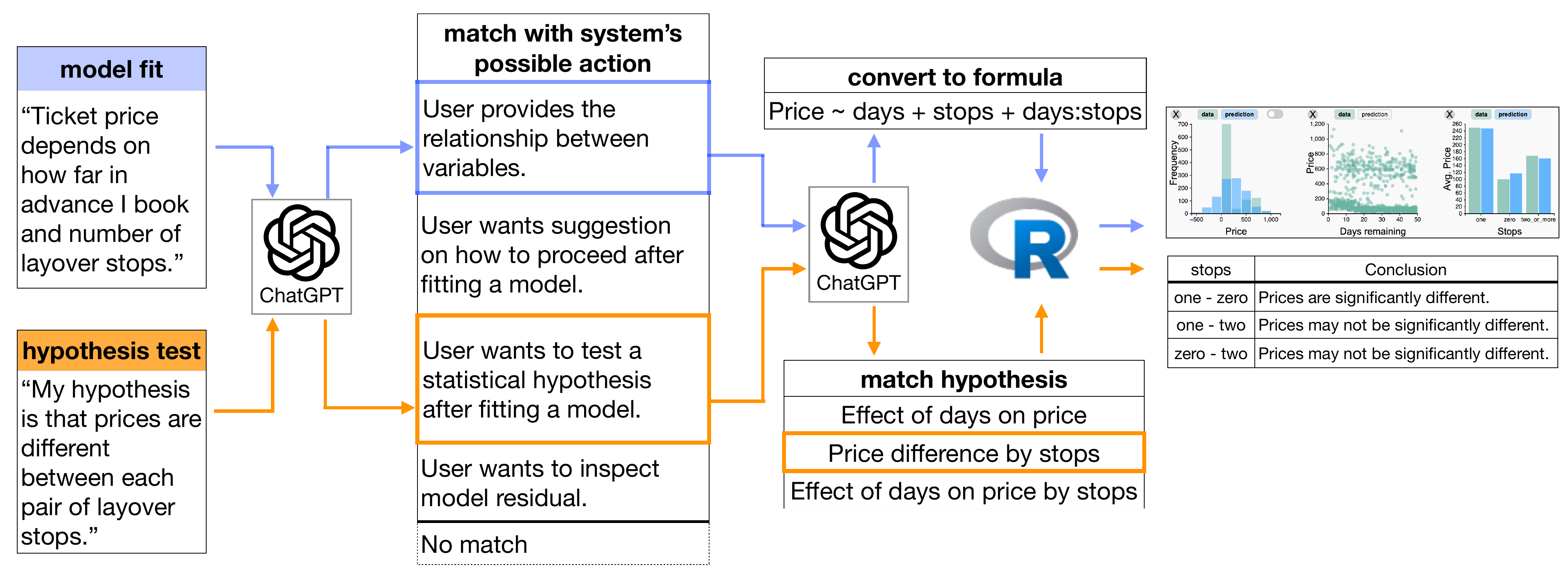}
    \caption{backend flow of the tool, with two example queries: one for fitting a model and one for testing a statistical hypothesis.}
    \label{fig:backend}
\end{figure*}

\section{Related Work}

\subsection{Natural language interface for visual data analysis}

Several works have explored user interaction with data through natural language. For instance, DataTone \cite{gao2015datatone} and Eviza \cite{setlur2016eviza} support querying and modifying visualizations via conversational input. A Wizard-of-Oz study by Choi et al. \cite{choi2019concept} investigated how users perform hypothesis-driven reasoning using natural language. With Large language models (LLMs) being increasingly integrated into data visualization systems, the gap between user intent and technical specification has been bridged. Multiple works have explored capabilities of LLMs for data analysis and visualization. For instance, Inala et al. \cite{inala2024data} explores how large language and multimodal models can enhance data analysis workflows by translating high-level user intentions into executable code, charts, and insights. Hong et al. \cite{hong2023conversational} created an LLM-driven interface capable of generating both visual and textual responses based on user utterances. Xu et al. \cite{xu2024exploring} studied the potential of using LLM to perform low-level visual analytic tasks. By allowing users to describe analyses, patterns, or hypotheses in natural language, such conversational interface can generate code, help interpret results, and suggest a plan of action. This reduces the need for users to possess the statistical and computational rigor, enhancing more accessible analysis flow. However, integrating LLMs in such workflows raises challenges related to ambiguity, correctness, and trust. 

Our work builds on this foundation by leveraging LLMs for high-level scaffolding such as translating user queries into structured commands while delegating rendering and computation to deterministic engines. This hybrid approach leverages the expressive flexibility of LLMs while maintaining the accuracy and transparency critical for visual data analysis.

\subsection{Visualization for statistical modeling}

Visualization has long played a role in helping users construct, refine, and interpret statistical models. Systems like ModelTracker \cite{amershi2015modeltracker} and Prospector \cite{krause2016interacting} allow users to interactively explore model parameters, residuals, and performance metrics, often with support for feature selection and debugging. EVM \cite{kale2023evm} enables users to incorporate model checks during exploratory visual analysis, and VMC \cite{guo2024vmc} implements the model check process as an R package, utilizing the ggdist package \cite{kay2023ggdist} for its core visualization capabilities. These tools typically assume structured interactions (e.g., sliders, dropdowns, variable selection) and require some modeling literacy. In contrast, our system initiates the modeling process via natural language, relying on generative AI to translate user intent into a formal model specification, while preserving traditional visual outputs for transparency, validation, and iterative refinement.

\section{Design Goals}

Our main goal is to design an approach for integrating GenAI as a supportive visual analysis tool, ensuring that the primary decision-making and analytical control remains with the user, while the AI serves strictly as an assistive aid.

\textbf{D1: Leverage GenAI to lower barriers to statistical modeling}: We integrate GenAI to make statistical modeling more accessible to users with limited statistical expertise. Users express analytic intent in natural language, and GenAI translates these queries into formal statistical specifications that are executed in the R backend.

\textbf{D2: Combine visualization and GenAI to support iterative, user-driven exploration}: Users explore models, inspect residuals, and refine hypotheses through interactive visualizations. We integrate GenAI to complement this process. When users identify patterns or mismatches in the visualizations, they can express revised intent in natural language, and GenAI translates these refinements into updated model specifications.


Overall, our system combines natural language interaction, statistical modeling, and interactive visualization. We consider how users can move fluidly between expressing analytic intent, specifying models, interpreting results, and refining hypotheses. While we use LLM to translate user intent into statistical model formulations, we intentionally constrain its role to high-level interpretation. The language model is not involved in core analytical tasks such as model fitting, diagnostics, or statistical testing. These steps are delegated entirely to the R backend. By positioning our tool between fully manual tools and fully automated assistants, we aim to offer a more trustworthy, human-centered approach to data analysis. Users retain agency and interpretive control, but benefit from a more natural and accessible interface for model specification and refinement. This design encourages a more iterative and exploratory approach to analysis, where users can quickly test hypotheses, visualize results, and adjust their thinking in response to feedback while staying grounded in statistically sound practices.

\section{Implementation}

In the following sections, we describe the implementation of our system. We use the GPT-4 model through OpenAI's API to facilitate interaction between the user and the tool. The LLM serves two primary purposes in our design. First, it helps interpret user input to guide the system in generating appropriate responses. Second, it acts as an initial translation layer that converts user intent into R formulation, which is then passed to the statistical backend. After detailing this high-level integration of LLM, we then describe how the statistical computations are performed on the backend. Finally, we describe the interactive visualization functionalities offered by the tool.

\subsection{Use of LLM for natural language interpretation}

\subsubsection{Guiding the system toward correct response}

Figure \ref{fig:backend} illustrates the backend process, specifically how user queries pass through the system and return results to the interface. Users start interacting with the system by entering queries in natural language. Depending on their current stage in the analysis process, they may ask the system to perform a variety of tasks. To accommodate this variability, we first define a set of supported tasks on the backend. Each task is represented by a Task Description that outlines its purpose and is paired with a corresponding executable function. When a user submits a request, we use LLM to match the freeform input to the most appropriate Task Description. This approach allows us to constrain the user's intent to a predefined set of valid actions, ensuring that only supported operations are executed. For example, if a user enters the query, ``Ticket price depends on how far in advance I book and the number of layover stops,'' they are describing a relationship between variables rather than expressing an intent to test a statistical hypothesis (e.g., by mentioning terms like ``hypothesis'' or ``significant result''). In this case, the system would match the query to the action ``User provides the relationship between variables'' and proceed accordingly, as illustrated in Figure \ref{fig:backend}. If the input does not match any task in the list, the system responds with: ``Please try a different query.''



\subsubsection{Model translation before statistical computation}

Beyond matching user intent to predefined system tasks, our tool also leverages GenAI to translate freeform natural language into structured model specifications and hypothesis tests. For instance, a user might input a request such as ``Fit a linear model predicting flight price from flight duration'' or ``Test whether the number of layover stops affects the price of tickets.'' These inputs are passed to the ChatGPT API, which is prompt-engineered to interpret the user's analytical goals and generate formal statistical representations suitable for execution in R.

For modeling tasks, user input is translated into an R-style formula, such as $price \sim duration$, which can be directly processed by model-fitting functions in the backend. For hypothesis testing, the system interprets contrasts or comparisons implied in the user's language and translates them into structured test specifications. For example, a statement like ``My hypothesis is that prices are different between each layover stop'' is interpreted as a test of average price differences between all pairs within the group.

These translations are handled exclusively by the language model in a constrained prompt environment, which ensures consistency in the output format. Once the structured model or hypothesis is generated, it is passed to the R backend, where statistical analysis is performed using libraries such as gamlss and emmeans. The results, including model coefficients, residuals, p-values, and fitted values, are returned to the front end for visualization and interpretation.

By limiting the GenAI’s role to translation and not analysis, we maintain a clear boundary between intent interpretation and statistical computation. This ensures that users benefit from natural language interaction without compromising the correctness, reproducibility, or transparency of the analytical workflow.

\subsection{Interface and visualization}

Following natural language input and statistical analysis, the frontend interface displays the results to users. It offers a range of interactive features that support exploratory visualization. An overview of the interface is shown in Figure \ref{fig:teaser}. Users can select any variable in the dataset, either individually or in pairs. When one continuous variable is selected, a histogram is displayed. When one categorical variable is selected, a bar chart showing the counts for each category appears. If two continuous variables are selected, the tool generates a scatter plot. When a continuous variable is paired with a categorical variable, the tool displays an average bar chart that shows the mean of the continuous variable for each category. Alternatively, a vertical scatterplot can be used to display individual, non-aggregated data points. The system also supports brushing and linking across charts, allowing users to highlight and compare data points interactively.

\begin{figure}
    \centering
    \includegraphics[width=1\linewidth]{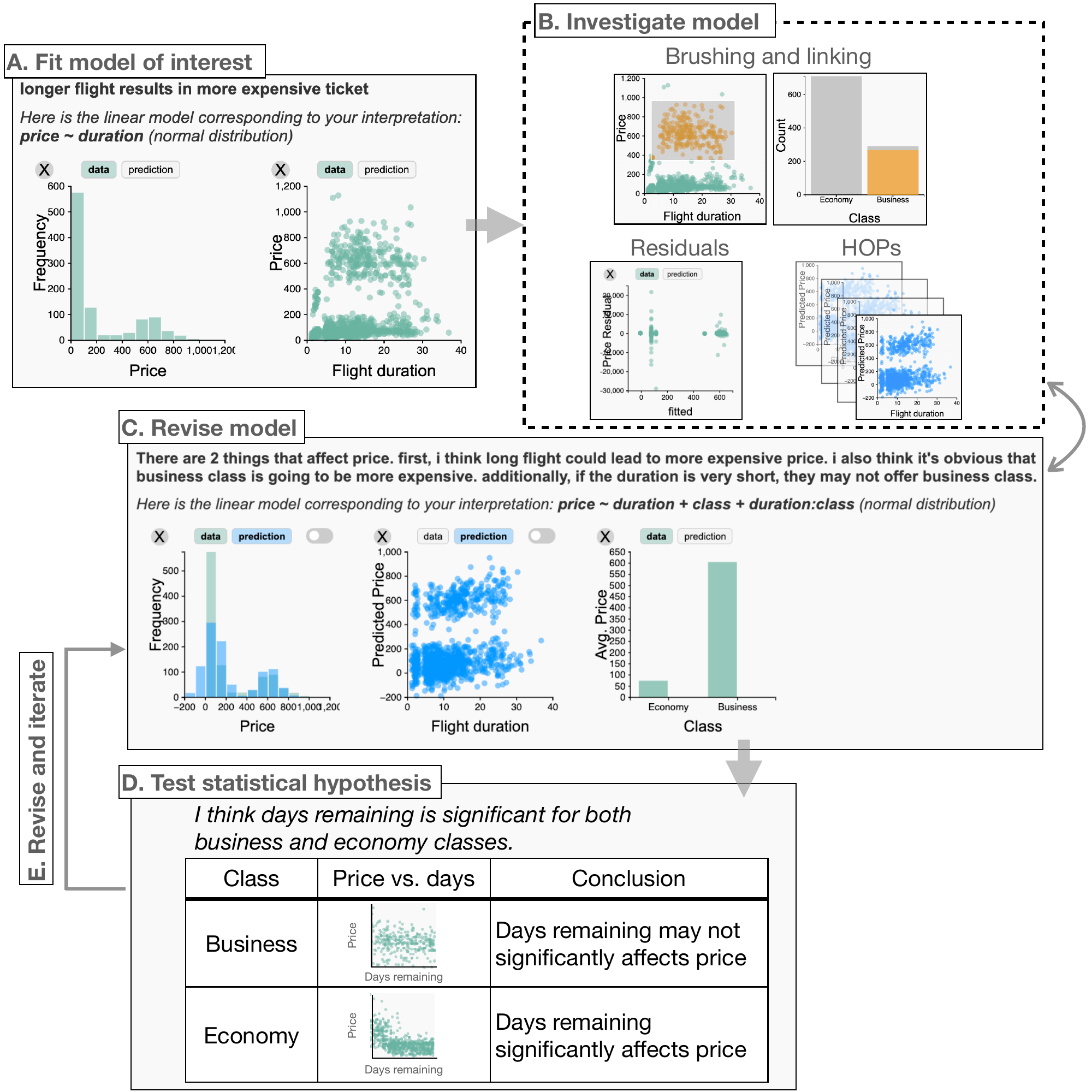}
    \caption{Iteration of the user workflow: The process begins with fitting the model through a query (A). Users can then optionally investigate and revise the model (B). This may involve exploring visualizations of other variables or inspecting model-related visualizations like residuals and Hypothetical Outcome Plots (HOPs). Users can revise the model (C) and also test a statistical hypothesis for a variable in the model (D). The entire process is likely iterative (E).}
    \label{fig:iteration}
\end{figure}

\section{Example Use-case}

In this section, we walk through an example use case to illustrate the potential of our tool. The process is also illustrated in Figure \ref{fig:iteration}. Amey, a business analyst at an airline company, is tasked with analyzing a flight price dataset to support a strategic business proposal aimed at increasing revenue. His primary goal is to identify which variables most strongly influence ticket prices, as understanding these factors allows the company to optimize pricing strategies, forecast revenue more accurately, and design targeted promotions. The airline's executive team expects Amey to deliver defensible evidence for pricing decisions that could influence millions in revenue, making accuracy and interpretability critical for his analysis. While Amey is proficient in creating ad hoc visualizations and dashboards, he has limited experience with formal statistical modeling. Without the tool, he would need to rely on dashboards and manual plotting to test his hypotheses. He would have to write R code to specify models, interpret diagnostics, and check statistical assumptions, which could be both time-consuming and error-prone given his background. To ensure his analysis remains rigorous and interpretable, Amey uses our tool to translate his questions into valid statistical formulations and generate interactive visual feedback that guides his exploration and insights.

Amey begins his analysis by confirming the simple assumption that longer flights are generally more expensive. He expresses this by entering, ``Longer flight results in a more expensive ticket,'' prompting the system to fit a model using flight duration as the sole predictor. The R backend fits the model, and the interface displays interactive visualizations, including predicted vs. observed prices and residual plots. Unlike static summaries, these visualizations allow Amey to immediately assess whether the model adequately explains price variation. Upon inspecting the plots, Amey notices a significant mismatch between predicted and actual prices. The residual plot indicates poor fit, and the scatterplot shows two distinct clusters of points. Using brushing and linking, Amey discovers that the clusters correspond to travel class (economy vs. business). Recognizing this as a key factor, Amey decides to refine the model and types: ``Include class as an additional variable.'' The backend refits the model and updates the visualizations. Predictions now align more closely with observed prices, and residuals appear more randomly distributed, indicating a better fit.

Encouraged by the improved fit but unsure how to proceed, Amey looks to the system for guidance. The system suggests possible next steps without taking control: ``You can inspect residuals, explore prediction variability using HOPs, or try a different distribution.'' Amey chooses to inspect the residuals, and the system reminds him that residuals should ideally display random patterns if the model is well-specified. Guided by this, Amey examines the visualization and notices non-random patterns, suggesting potential issues with the current model. He informs the system: ``The residual patterns don’t look random.'' In response, the system suggests an optional refinement: ``A skewed distribution may better capture these patterns. Would you like to try it?'' Amey agrees, and the backend refits the model using the suggested distribution. The updated residual plot shows a better fit, which Amey confirms visually. Through this process, visualization guides Amey’s reasoning, the system provides contextual guidance, and GenAI translates Amey’s intent into updated model specifications, all while leaving the decision-making under Amey’s control.

With a well-fitting model, Amey tests a specific hypothesis: ``Does flight duration affect price differently for economy and business class?'' GenAI translates this question into a formal hypothesis test, and the system returns both a summary table and an interactive visualization comparing slopes across classes. Amey observes that flight duration strongly predicts prices for economy tickets, while its effect is weaker and not statistically significant for business class. Based on these findings, the airline can adopt a segmented pricing strategy by aligning economy-class fares closely with flight duration while setting business-class prices using demand and value-based factors rather than distance alone.

\section{Discussion and Future Direction}

Our system demonstrates that GenAI can be effectively integrated into visual statistical modeling workflows without assuming control over analytical reasoning. In our design, we leverage an LLM to constrain user input to the scope of the tool and translate user intent into precise statistical formulations, while interactive visualizations surface model behavior, residual patterns, and hypothesis comparisons. By combining GenAI-assisted intent translation with visual feedback, users can iteratively refine models while maintaining agency over analytical decisions. This design reduces the risk of hallucinated results, preserves the interpretability of statistical outputs, and enables users to validate their reasoning through visualization. Such integration supports a safer, more transparent form of AI assistance within visual analytics workflows, which is particularly valuable in domains where rigor, reproducibility, and user accountability are essential.

A key trade-off in our design is the limited scope of automation. While this promotes interpretability, it may also constrain usability for users expecting more guidance on model selection, assumption checking, or interpretation. Future iterations of the system could include optional suggestions from GenAI such as surfacing alternative model structures or highlighting potential violations of modeling assumptions without relinquishing full control to the model.
We also envision expanding the system’s capabilities to support richer dialogue-based interactions, enabling users to iteratively refine hypotheses and models in greater detail through natural language. Additionally, future work could explore personalized prompting strategies that adapt the translation process to a user’s background, domain, or elicited beliefs \cite{koonchanok2021data, koonchanok2023visual, koonchanok2024trust}. Finally, conducting user studies with both novices and experts will be critical to evaluating how this design impacts understanding, trust, and decision quality compared to fully automated or fully manual alternatives.

\section{Conclusion}

This work presents a human-centered approach to integrating generative AI into visual statistical modeling workflows. Rather than relying on LLMs to conduct analysis or generate insights, our system uses GenAI in a narrowly scoped, high-level role: translating natural language intent into formal statistical formulations, while interactive visualizations surface model behavior, residual patterns, and hypothesis comparisons. This combination enables broader accessibility to modeling tools while preserving analytical rigor, transparency, and user agency. By situating our tool between fully manual and fully automated systems, we offer an alternative path that leverages the strengths of GenAI within visual analytics workflows without compromising methodological integrity. As generative technologies continue to evolve, we believe this hybrid, visualization-driven approach offers a promising direction for developing assistive and trustworthy tools in data science and beyond. Ultimately, our work highlights how carefully integrating GenAI with interactive visualization can support interpretable, collaborative, and iterative modeling while keeping analysts in control of critical decisions.

\bibliographystyle{abbrv-doi}

\bibliography{template}

\begin{thebibliography}{10}

\bibitem{amershi2015modeltracker}
S.~Amershi, M.~Chickering, S.~M. Drucker, B.~Lee, P.~Simard, and J.~Suh.
\newblock Modeltracker: Redesigning performance analysis tools for machine learning.
\newblock In {\em Proceedings of the 33rd annual ACM conference on human factors in computing systems}, pp. 337--346, 2015.

\bibitem{choi2019concept}
I.~K. Choi, T.~Childers, N.~K. Raveendranath, S.~Mishra, K.~Harris, and K.~Reda.
\newblock Concept-driven visual analytics: an exploratory study of model-and hypothesis-based reasoning with visualizations.
\newblock In {\em Proceedings of the 2019 chi conference on human factors in computing systems}, pp. 1--14, 2019.

\bibitem{gao2015datatone}
T.~Gao, M.~Dontcheva, E.~Adar, Z.~Liu, and K.~G. Karahalios.
\newblock Datatone: Managing ambiguity in natural language interfaces for data visualization.
\newblock In {\em Proceedings of the 28th annual acm symposium on user interface software \& technology}, pp. 489--500, 2015.

\bibitem{guo2024vmc}
Z.~Guo, A.~Kale, M.~Kay, and J.~Hullman.
\newblock Vmc: A grammar for visualizing statistical model checks.
\newblock {\em IEEE Transactions on Visualization and Computer Graphics}, 2024.

\bibitem{hong2023conversational}
M.-H. Hong and A.~Crisan.
\newblock Conversational ai threads for visualizing multidimensional datasets.
\newblock {\em arXiv preprint arXiv:2311.05590}, 2023.

\bibitem{inala2024data}
J.~P. Inala, C.~Wang, S.~Drucker, G.~Ramos, V.~Dibia, N.~Riche, D.~Brown, D.~Marshall, and J.~Gao.
\newblock Data analysis in the era of generative ai.
\newblock {\em arXiv preprint arXiv:2409.18475}, 2024.

\bibitem{kale2023evm}
A.~Kale, Z.~Guo, X.~L. Qiao, J.~Heer, and J.~Hullman.
\newblock Evm: Incorporating model checking into exploratory visual analysis.
\newblock {\em IEEE Transactions on Visualization and Computer Graphics}, 30(1):208--218, 2023.

\bibitem{kay2023ggdist}
M.~Kay.
\newblock ggdist: Visualizations of distributions and uncertainty in the grammar of graphics.
\newblock {\em IEEE Transactions on Visualization and Computer Graphics}, 30(1):414--424, 2023.

\bibitem{koonchanok2021data}
R.~Koonchanok, P.~Baser, A.~Sikharam, N.~K. Raveendranath, and K.~Reda.
\newblock Data prophecy: Exploring the effects of belief elicitation in visual analytics.
\newblock In {\em Proceedings of the 2021 CHI Conference on Human Factors in Computing Systems}, pp. 1--12, 2021.

\bibitem{koonchanok2024trust}
R.~Koonchanok, M.~E. Papka, and K.~Reda.
\newblock Trust your gut: Comparing human and machine inference from noisy visualizations.
\newblock {\em IEEE Transactions on Visualization and Computer Graphics}, 2024.

\bibitem{koonchanok2023visual}
R.~Koonchanok, G.~Y. Tawde, G.~R. Narayanasamy, S.~Walimbe, and K.~Reda.
\newblock Visual belief elicitation reduces the incidence of false discovery.
\newblock In {\em Proceedings of the 2023 CHI conference on human factors in computing systems}, pp. 1--17, 2023.

\bibitem{krause2016interacting}
J.~Krause, A.~Perer, and K.~Ng.
\newblock Interacting with predictions: Visual inspection of black-box machine learning models.
\newblock In {\em Proceedings of the 2016 CHI conference on human factors in computing systems}, pp. 5686--5697, 2016.

\bibitem{morris2024prompting}
M.~R. Morris.
\newblock Prompting considered harmful.
\newblock {\em Communications of the ACM}, 67(12):28--30, 2024.

\bibitem{setlur2016eviza}
V.~Setlur, S.~E. Battersby, M.~Tory, R.~Gossweiler, and A.~X. Chang.
\newblock Eviza: A natural language interface for visual analysis.
\newblock In {\em Proceedings of the 29th annual symposium on user interface software and technology}, pp. 365--377, 2016.

\bibitem{xu2024exploring}
Z.~Xu and E.~Wall.
\newblock Exploring the capability of llms in performing low-level visual analytic tasks on svg data visualizations.
\newblock In {\em 2024 IEEE Visualization and Visual Analytics (VIS)}, pp. 126--130. IEEE, 2024.

\end{thebibliography}
\end{document}